\documentclass[a4paper,12pt]{article}
\usepackage[left=2cm,right=2cm, top=2cm,bottom=2cm,bindingoffset=0cm]{geometry}
\usepackage[T2A]{fontenc}
\usepackage{graphicx}
\usepackage{cite}
\usepackage{amssymb,amsmath,latexsym}
\usepackage{caption}
\usepackage{authblk}
\usepackage{color}
\usepackage{amssymb}
\usepackage{amsfonts}
\usepackage{bigints}

\usepackage{epsfig}

\newtheorem{theorem}{Theorem}
 \newtheorem{corollary}{Corollary}
\newtheorem{proposition}{Proposition}
\newtheorem{remark}{Remark}
\numberwithin{equation}{section} \numberwithin{lemma}{section} \numberwithin{theorem}{section} \numberwithin{corollary}{section} \numberwithin{proposition}{section}
\numberwithin{remark}{section}

\begin{document}

\title{On linearizability via nonlocal transformations and first integrals for second-order ordinary differential equations}
\author{Dmitry I. Sinelshchikov}

\affil{National Research University Higher School of Economics, Moscow, Russia}

\maketitle

\begin{abstract}
Nonlinear second-order ordinary differential equations are common in various fields of science, such as physics, mechanics and biology. Here we provide a new family of integrable second-order ordinary differential equations by considering the general case of a linearization problem via certain nonlocal transformations. In addition, we show that each equation from the linearizable family admits a transcendental first integral and study particular cases when this first integral is autonomous or rational. Thus, as a byproduct of solving this linearization problem we obtain a classification of second-order differential equations admitting a certain transcendental first integral. To demonstrate effectiveness of our approach, we consider several examples of autonomous and non-autonomous second order differential equations, including generalizations of the Duffing and Van der Pol oscillators, and construct their first integrals and general solutions. We also show that the corresponding first integrals can be used for finding periodic solutions, including limit cycles, of the considered equations.
\end{abstract}

\section{Introduction}

Here we consider the following family of non-autonomous nonlinear second-order differential equations
\begin{equation}\label{eq:eq1}
  y_{zz}+f(z,y) y_{z}^{2}+g(z,y)y_{z}+h(z,y)=0,
\end{equation}
where $f$, $g$ and $h$ are arbitrary sufficiently smooth functions. We assume that $gh\not\equiv0$ and that $f^{2}+g_{y}^{2}+h_{yy}^{2}\neq0$, i.e. we exclude the linear subcase of \eqref{eq:eq1} from the consideration.

Equations from family \eqref{eq:eq1} often appear in numerous applications in mechanics, physics and so on \cite{Guckenheimer1983,Andronov}. Therefore, various aspects of integrability of \eqref{eq:eq1} have been studied in a number of works (see, e.g., \cite{Duarte,Meleshko2010,Muriel2009,Muriel2010,Muriel2011,Muriel2011a,Meleshko2012,Muriel2018,Nucci2010,Nucci2010a,Bagderina2013,Bagderina2016,Guha2019,Gine2019a,Gine2019b}). For example, in \cite{Muriel2009,Muriel2011,Meleshko2012,Muriel2018} authors considered applications of several linearizing transformations and $\lambda$-symmetries for finding first integrals of equations from family \eqref{eq:eq1}. In particular, in \cite{Muriel2009} it was shown that equations admitting a linear with respect to the first derivative first integral form exactly the same class as equations linearized to the Laguerre normal form of linear second order differential equations (the latter class was obtained in \cite{Duarte}). Authors of \cite{Muriel2009} also demonstrated that equations from the corresponding class possess a certain $\lambda$-symmetry and there is a subclass of completely integrable equations with two independent first integrals. In \cite{Muriel2010,Muriel2011,Muriel2011a,Meleshko2012} various connections between linearizability of second-order differential equations and the existence of certain first integrals, in particular rational ones, were studied. Authors of \cite{Nucci2010,Nucci2010a} applied the Jacobi last multiplier approach for studying integrability of \eqref{eq:eq1}, while in \cite{Bagderina2013,Bagderina2016} equivalence problems via point transformations were studied. Connections via nonlocal transformations between equations from \eqref{eq:eq1} and various Painlev\'e type equations were considered in \cite{Kudryashov2017a,Sinelshchikov2017,Sinelshchikov2018,Sinelshchikov2019}.

Here we deal with the linearization problem for \eqref{eq:eq1} via the generalized Sundman transformations, which have the form
\begin{equation}\label{eq:eq2}
w=F(z,y), \quad d\zeta=G(z,y)dz,
\end{equation}
where $F$ and $G$ are some sufficiently smooth functions satisfying $F_{y}G\neq0$. This problem was previously studied in \cite{Duarte,Meleshko2010}. While in \cite{Duarte} linearization to the Laguerre normal form of second order linear differential equation, namely to the equation $w_{\zeta\zeta}=0$, was considered, in \cite{Meleshko2010} it was shown that for the linearization via transformations \eqref{eq:eq2} it is insufficient to use the Laguerre normal form of a linear second order differential equation and connections between \eqref{eq:eq1} and
\begin{equation}\label{eq:eq5}
w_{\zeta\zeta}+\beta w_{\zeta}+\alpha w=0,
\end{equation}
were considered. Here $\alpha\neq0,\beta\neq0$ are arbitrary parameters. Although authors of \cite{Meleshko2010} studied the equivalence problem between \eqref{eq:eq1} and \eqref{eq:eq5} via \eqref{eq:eq2}, only a particular case of transformations \eqref{eq:eq2}, specifically the case of $F_{z}=0$, was considered. However, it is known that there are interesting from an applied point of view nonlinear oscillators that can be linearized via \eqref{eq:eq2} only if $F_{z}\neq0$ (see, e.g. \cite{Demina2019}).  Therefore, in this work we consider the full linearization problem for \eqref{eq:eq1} and find all equations from family \eqref{eq:eq1} that can be linearized with the help of \eqref{eq:eq2} with $F_{z}\neq0$. We demonstrate that there are nontrivial examples of equations from \eqref{eq:eq1} that can be linearized only via \eqref{eq:eq2} with $F_{z}\neq0$. Furthermore, we show that each linearizable equation from \eqref{eq:eq1} admits a certain first integral, which can be explicitly constructed via the parameters of the studied equation and linearizing transformations. This follows from the fact that linear equation \eqref{eq:eq5} possesses an autonomous first integral and we believe that this is the first time when the corresponding first integrals are obtained for linearizable equations from \eqref{eq:eq1}. We also separately consider the cases when this first integral is autonomous or a rational/polynomial function. Finally, let us remark that authors of \cite{Meleshko2010} also included a constant parameter $\gamma$ in \eqref{eq:eq5}, but it can be easily removed via the transformation $F\rightarrow F+\gamma/\alpha$, and, consequently, we do not take it into consideration.

Notice also that the linearization problem for family \eqref{eq:eq1} via a more general class of nonlocal transformations, when the function $G$ in \eqref{eq:eq2} depends on $y_{z}$ was considered (see, e.g. \cite{Muriel2011,Muriel2011a,Chandrasekar2006} and references therein). For instance, in \cite{Muriel2011a} linearization problem for \eqref{eq:eq1} via \eqref{eq:eq2} with $G(z,y,y_{z})=G_{1}(z,y)y_{z}+G_{2}(z,y)$ was studied. Authors of \cite{Muriel2011a} showed that equations for this linearizable class possess a certain rational first integral and a $\lambda$-symmetry, which can be calculated in terms of the coefficients of the corresponding equation.

The rest of this work is organized as follows. In the next Section we present the equivalence criterion for \eqref{eq:eq1} and \eqref{eq:eq5}. We also show how to construct a first integral for linearizable equation from \eqref{eq:eq1} and present several interesting subcases of linearizable equations from \eqref{eq:eq1}, namely Darboux integrable cases and equations with rational non-autonomous first integrals. In Section 3 we provide several examples of linearizable equations from \eqref{eq:eq1} including parametrically forced generalizations of the Duffing and Van der Pol equations. In the last section we briefly discuss and summarize our results.

\section{New integrability conditions}

Let us start with some preliminary results. First we introduce a canonical form of \eqref{eq:eq1} with respect to \eqref{eq:eq2}.

\begin{proposition}
Family of equations \eqref{eq:eq1} is closed with respect to \eqref{eq:eq2} and its canonical form is
\begin{equation}\label{eq:eq1a}
  y_{zz}+g(z,y)y_{z}+h(z,y)=0.
\end{equation}
\end{proposition}
\textit{Proof.}
The closedness of \eqref{eq:eq1} with respect to \eqref{eq:eq2} can be checked by direct calculations. Thus, without loss of generality, one can assume that $f(z,y)=0$. Indeed, substituting the transformation $\tilde{y}=\int \exp\{\mathfrak{f}\}dy$, which is a particular case of \eqref{eq:eq2}, into \eqref{eq:eq1} we get
\begin{equation}
\tilde{y}_{zz}+\tilde{g}(z,\tilde{y})\tilde{y}_{z}+\tilde{h}(z,\tilde{y})=0,
\label{eq:eq3}
\end{equation}
where
\begin{equation}
\begin{gathered}
\mathfrak{f}=\int f dy, \quad \tilde{g}=g-2\mathfrak{f}_{z},\\
\tilde{h}={\rm e}^{\mathfrak{f}}h+(2\mathfrak{f}_{z}-g)\int \mathfrak{f}_{z} {\rm e}^\mathfrak{f}dy -\int  ( \mathfrak{f}_{zz}+\mathfrak{f}_{z}^{2}){\rm e}^{ \mathfrak{f}} dy.
\label{eq:eq3a}
\end{gathered}
\end{equation}
In order to obtain results for family of equations \eqref{eq:eq1} from the results for \eqref{eq:eq3} we need to make the following substitutions
\begin{equation}
\begin{gathered}
y\rightarrow \int {\rm e}^{\mathfrak{f}}dy, \quad g\rightarrow g+2\mathfrak{f}_{z}, \\
h\rightarrow {\rm e}^{-\mathfrak{f}} \left(h+\int \mathfrak{f}_{z} {\rm e}^\mathfrak{f}dy g +\int  ( \mathfrak{f}_{zz}+\mathfrak{f}_{z}^{2}){\rm e}^{ \mathfrak{f}} dy\right).
\label{eq:eq3b}
\end{gathered}
\end{equation}
This completes the proof. $\Box$

Consequently, further we assume that $f(z,y)=0$ and study equivalence problem for \eqref{eq:eq1a}.

Now let us show that \eqref{eq:eq5}  has an autonomous first integral that can be used for constructing first integrals for linearizable equations from \eqref{eq:eq1a}. Indeed, it is easy to verify that the following expression
\begin{equation}\label{eq:eq6a}
I=\left(2w_{\zeta}+(\beta+\rho)w\right)^{\rho+\beta}\left(2w_{\zeta}+(\beta-\rho)w\right)^{\rho-\beta},
\end{equation}
where $\rho=\sqrt{\beta^{2}-4\alpha}\neq0$, is a first integral of \eqref{eq:eq5}. If $\rho=0$, instead of \eqref{eq:eq6a} one needs to use
\begin{equation}\label{eq:eq6b}
I=(2w_{\zeta}+\beta w)\exp\left\{\frac{\beta w}{2w_{\zeta}+\beta w}\right\}.
\end{equation}
Notice also that if $\rho$ is imaginary, i.e. $\beta^{2}-4\alpha<0$,  first integral \eqref{eq:eq6a} can be transformed into a real form as follows
\begin{equation}
I=\ln\left\{\alpha w^{2}+\beta w w_{\zeta}+w_{\zeta}^{2}\right\}-\frac{2\beta}{\sqrt{-\rho^{2}}}\arctan\left\{\frac{2w_{\zeta}+\beta w}{\sqrt{-\rho^{2}}w}\right\}.
\label{eq:eq6c}
\end{equation}

\begin{remark}
Notice that from the results of \cite{Muriel2009} it follows that \eqref{eq:eq5} possesses two functionally independent first integrals, which are linear functions with respect to $w_{\zeta}$. These integrals are
  \begin{equation}\label{eq:eq6d}
    I_{1}={\rm e}^{\frac{\beta+\rho}{2}\zeta}(2w_{\zeta}+(\beta-\rho)w), \quad  I_{2}={\rm e}^{\frac{\beta-\rho}{2}\zeta}(2w_{\zeta}+(\beta+\rho)w).
  \end{equation}
However, \eqref{eq:eq6d} cannot be used for constructing first integrals of linearizable equations from \eqref{eq:eq1a}, since transformations \eqref{eq:eq2} do not map a non-autonomous first integral of \eqref{eq:eq5} into a first integral of a linearizable equation from \eqref{eq:eq1a}.

On the other hand, integral \eqref{eq:eq6a} can be easily obtained from \eqref{eq:eq6d} as $I=I_{1}^{\rho-\beta}I_{2}^{\rho+\beta}$. If one considers another function of $I_{1}$ and $I_{2}$ that gives an autonomous first integral of \eqref{eq:eq5}, one obtains an integral, that is a function of \eqref{eq:eq6a}, since \eqref{eq:eq5} can admit at most one, up to a functional dependence, autonomous first integral. In other words, any autonomous first integral of \eqref{eq:eq5} is a function of \eqref{eq:eq6a}. Thus, only \eqref{eq:eq6a} (or any function of it) can be used for constructing first integrals for linearizable equations from \eqref{eq:eq1a}.  The cases of $\beta^{2}-4\alpha<0$ and $\beta^{2}-4\alpha=0$ can be treated in a similar way.

\end{remark}

Let us proceed with the main result of this section and obtain the necessary and sufficient conditions for \eqref{eq:eq1a} to be equivalent to \eqref{eq:eq5} via \eqref{eq:eq2}. Necessary conditions can be obtained if one substitutes the expressions for $w$, $w_{\zeta}$ and $w_{\zeta\zeta}$ via $F$ and $G$ into \eqref{eq:eq5}. This yields to
\begin{equation}
\label{eq:eq7}
y_{zz}+g y_{z}+h=0,
\end{equation}
provided that
\begin{equation}
\label{eq:eq7a}
GF_{yy}-F_{y}G_{y}=0,
\end{equation}
holds. Here
\begin{equation}
\label{eq:eq7b}
g=\frac{2GF_{yz}-F_{y}G_{z}-F_{z}G_{y}+\beta G^{2}F_{y}}{GF_{y}}, \quad h=\frac{GF_{zz}-F_{z}G_{z}+\beta G^{2}F_{z}+\alpha G^{3}F}{G F_{y}}.
\end{equation}
Therefore, equation \eqref{eq:eq1a} can be transformed into \eqref{eq:eq5} if it is of the form \eqref{eq:eq7} and \eqref{eq:eq7a} holds.

Conversely, if the functions $F$ and $G$ satisfy \eqref{eq:eq7a}, \eqref{eq:eq7b} then equation can be transformed into \eqref{eq:eq5} with the help of \eqref{eq:eq2}. As a result, compatibility conditions for the following overdetermined system of partial differential equations for the functions $F$ and $G$
\begin{equation}
\begin{gathered}
\label{eq:eq8}
GF_{yy}-G_{y}F_{y}=0, \\
gGF_{y}-2GF_{yz}+F_{y}G_{z}+F_{z}G_{y}-\beta G^{2}F_{y}=0,\\
h G F_{y}+G_{z}F_{z}-G F_{zz}-\alpha G^{3}F-\beta G^{2} F_{z}=0.
\end{gathered}
\end{equation}
give us the necessary and sufficient conditions for \eqref{eq:eq1a} to be equivalent to \eqref{eq:eq5} via \eqref{eq:eq2}.

Now our goal is to explicitly find correlations on the functions $f$ and $g$ that provide compatibility of \eqref{eq:eq8} and, hence, define equations of the form \eqref{eq:eq1a} that can be both linearizable and admit a certain first integral. Although direct computation of the compatibility conditions for \eqref{eq:eq8} is quite cumbersome, we can considerably simplify this system, which allows us to explicitly find required compatibility conditions.

Solving the first equation from \eqref{eq:eq8} we get that
\begin{equation}
G=AF_{y},
\label{eq:eq9}
\end{equation}
where $A=A(z)\not\equiv 0$ is an arbitrary sufficiently smooth function. With the help of this relation, from \eqref{eq:eq8} we obtain
\begin{equation}
\begin{gathered}
\beta A F_{y}-g-\frac{A_{z}}{A}-\frac{F_{z}F_{yy}}{F_{y}^{2}}+\frac{F_{yz}}{F_{y}}=0,\vspace{0.1cm}\\
\alpha A^{2}FF_{y}+\beta A F_{z}-h-\left(\frac{A_{z}}{A}+\frac{F_{yz}}{F_{y}}\right)\frac{F_{z}}{F_{y}}+\frac{F_{zz}}{F_{y}}=0.
\label{eq:eq10}
\end{gathered}
\end{equation}
The first equation from \eqref{eq:eq10} can be integrated once with respect to $y$. As a result, we obtain
\begin{equation}
F_{z}+\left(\beta A F -m -C-\frac{A_{z}}{A}y\right)F_{y}=0,
\label{eq:eq10a}
\end{equation}
where $m_{y}=g$ and $C(z)$ is an arbitrary sufficiently smooth function.

With the help of \eqref{eq:eq10a}, from the second equation from \eqref{eq:eq10} we get
\begin{equation}
FF_{y}+\frac{1}{\alpha A^{2}}\left(m_{z}+C_{z}-h+\frac{A_{zz}}{A}y-\frac{A_{z}}{A}m+\frac{A_{z}}{A}C-2\frac{A_{z}^{2}}{A^{2}}y\right)=0.
\label{eq:eq10b}
\end{equation}
Introducing in \eqref{eq:eq10a}, \eqref{eq:eq10b} the following notations
\begin{equation}
p=m+C+\frac{A_{z}}{A}y, \quad l=\frac{\beta^{2}}{\alpha A^{2}}\left(\frac{A_{z}}{A}p+h-p_{z}\right), \quad L=\beta F,
\label{eq:eq10c}
\end{equation}
we have
\begin{equation}
\begin{gathered}
LL_{y}-l=0, \quad LL_{z}+A l L-lp=0.
\label{eq:eq11}
\end{gathered}
\end{equation}
This system is quite simple in comparison with \eqref{eq:eq8}. If we consider \eqref{eq:eq11} as an overdetermined system for the function $L$, the corresponding compatibility conditions give the necessary and sufficient conditions for linearization of \eqref{eq:eq1a} via \eqref{eq:eq2} provided that one takes into account notations \eqref{eq:eq10c}.

The compatibility conditions for \eqref{eq:eq11} split into four separate cases: the generic case and three particular cases. During the computation of the compatibility conditions we assume that $L\neq0$, $l\neq0$ and $A\neq0$ since otherwise transformations \eqref{eq:eq2} degenerate. To simplify further representation we introduce the following notations
\begin{equation}
P=l_{yy}, \quad Q=p_{yy}, \quad R=pl_{y}+lp_{y}-l_{z}.
\end{equation}

The generic case of the compatibility conditions is
\begin{equation}
\begin{gathered}
A^{2}Pl^{3}R_{y}R-P^{2}A^{4}l^{5}+ \left( 8\,A^{4}l^{4}l_{y}^{2}-7\,A^{2}R^{2}l^{2}l_{y}+R^{4} \right)P-\\
-A^{2}l_{y}l^{3}{R_{y}}^{2}+l_{y}R \left( 6 A^{2}l^{2}l_{y}-R^{2} \right) R_{y}-A^{2}l_{y}^{3}l \left( 16\,A^{2}l^{2}l_{y}-3\,R^{2} \right)=0, \vspace{0.2cm} \\
A^{3}PQl^{4}+ \left(2AR^{2}lp_{y}-A^{3}Rl^{2}l_{y}-A^{3}l^{3}R_{y}+ARlpR_{y}-AR^{3}-\right. \\ \left.-ARlR_{z}+R^{2}lA_{z}\right) P
-Al_{y}l \left( 4\,A^{2}l^{2}l_{y}-R^{2} \right) Q-l_{y} \left( l_{y}R-lR_{y} \right)  \left( 6\,A^{3}ll_{y} -\right. \\ \left.-2ARp_{y}-ApR_{y}+AR_{{z}}-RA_{z} \right)=0,\vspace{0.2cm} \\
l \left( 5A^{2}Rl^{2}l_{y}-A^{2}l^{3}R_{y}-R^{3} \right) P_{y}-7A^{2}P^{2}l^{3}R+A^{2}Pl^{4}R_{yy}-ll_{y} \left( 4A^{2}l^{2}l_{y}-R^{2}\right) R_{yy}-\\-
 \left( 4A^{2}l^{2}l_{y}+3R^{2} \right)  \left( Rl_{y}-lR_{y} \right) P-3l_{y} \left( Rl_{y}-lR_{y} \right)\left( 4A^{2}ll_{y}^{2}-RR_{y} \right)=0,
\label{eq:cc_case1}
\end{gathered}
\end{equation}
while the function $L$ is given by
\begin{equation}
L=\frac {Al \left(lR_{y}l_{y}- PRl-Rl_{y}^{2} \right) }{A^{2}P{l}^{2}l_{y}-4\,A^{2}ll_{y}^{3}-PR^{2}+RR_{y}l_{y}}\,\,\,.
\label{eq:cc_case1_L}
\end{equation}

Let us briefly describe the process of computation of conditions \eqref{eq:cc_case1} and expression \eqref{eq:cc_case1_L}. We consider \eqref{eq:eq11} as an overdetermined system of equations for $L$ and apply the Riquier--Janet compatibility theory (see, e.g. \cite{Reid1996}) for  computing the corresponding compatibility conditions. This is done via calculating various mixed partial derivatives of $L$ with respect to $z$ and $y$ and comparing them. The comparison of $L_{yz}$ and $L_{zy}$ leads to the expression for $L^{2}$ via $p$ and $l$. Then, with the help of this expression and expressions for $Lzyy$, $Lyyz$ and $Lyzy$ we find expression \eqref{eq:cc_case1_L} and the first condition from \eqref{eq:cc_case1}. Further computing and comparing third order mixed derivatives of $L$ we obtain the second condition from \eqref{eq:cc_case1}. Finally, with the help of expressions for $L_{yyyz}$, $L_{zyyy}$ and $L_{yyzy}$ we find the last compatibility condition from \eqref{eq:cc_case1}. Computation of further mixed partial derivatives of $L$ does not lead to new compatibility conditions. In addition, to verify that all compatibility conditions are obtained we compare our results with those produced by the Rif package \cite{Reid1996}. Our results and results produced by Rif coincide. Let us remark that further we do not provide details of the computation of the compatibility conditions since they are similar to those given above.

Now we need to consider particular cases of the compatibility conditions. First, we deal with the case when the denominator of \eqref{eq:cc_case1_L} vanishes. As a result, we get the following relations
 \begin{equation}
\begin{gathered}
A^{2}Pl^{3}-4\,A^{2}l^{2}l_{y}^{2}+R^{2}l_{y}=0,\\
l^{2} \left( Plp-4\,pl_{y}^{2}+5\,Rl_{y}-lR_{y} \right) A^{2}-R^{2} \left( R-pl_{y} \right)=0,\\
  \left( APp+AQl+2\,Al_{y}p_{y}-AR_{y}+A_{z}l_{y} \right) L^{4}-R_{z}L^{3}+\\
  +l \left( 2\,Alp_{y}+2\,Apl_{y}-3\,AR+lA_{z} \right) L^{2}+l \left( 2\,A^{2}l^{2}+Rp \right) L-2\,Apl^{3}=0,\\
4\,l^{2}RA^{2} \left( A^{2}l^{2}+5\,Rp \right) l_{y}^{2}+5\,Ql_{y}A^{4}l^{6}-l^{4}R^{2}QA^{2}-A^{4}l^{7}Q_{y}+2\,R^{5}-\\
-2\,R \left( -2\,A^{2}Rl^{3}p_{y}+A^{2}l^{3}pR_{y}+6\,A^{2}R^{2}l^{2}+A^{2}l^{3}R_{z}-ARl^{3}A_{z}+2\,R^{3}p \right) l_{y}=0,\\
\left( 2\,Al_{zzy}+4\,A_{z}l_{{z,y}}+2\,l_{y}A_{zz}\right) L^{4}+ \left(2 l_{zzz}-A^{2}Plp-A^{2}p l_{y}^{2}-2\,lp_{zzy}-\right. \\ \left.
-2\,pl_{zzy}-2\,l_{y}p_{zz}-4\,l_{z}p_{zy}-4\,p_{z}l_{zy}-2\,p_{y}l_{zz}\right)L^{3}+\left(APl{p}^{2}+AQl^{2}p+3Alpl_{y}p_{y}+\right. \\ \left. +A{p}^{2}l_{y}^{2}-4\,Al^{2}p_{zy}+5\,Alpl_{zy}-2\,All_{y}p_{z}-6\,All_{z}p_{y}-Apl_{y}l_{z}-\right.\\
\left.-2\,l^{2}A_{z}p_{y}+8\,lpA_{z}l_{y}+8\,All_{zz}+6\,A{l_{z}}^{2}+2l^{2}A_{zz}+10lA_{z}l_{z}\right) L^{2}-\\
 -l \left( 4 A^{2}l^{2}p_{y}+A^{2}lpl_{y}-24A^{2}ll_{z}-12Al^{2}A_{z}+6lpp_{zy}+6 p^{2}l_{zy}+6 pl_{y}p_{z}+\right. \\ \left.
 +6pl_{z}p_{y}-6pl_{zz} \right) L-l^{2} \left(7\,Alpp_{y}-12\,A^{3}l^{2}+3Ap^{2}l_{y}+2Alp_{z}-9Apl_{z}-2lpA_{z} \right)=0,
\label{eq:cc_case2}
\end{gathered}
\end{equation}
and $L$ is given by
 \begin{equation}
\begin{gathered}
L=Al^{2} \Big( 2\,A^{2}l^{2} \left( A^{2}l^{2}-5\,Rp\right) l_{y}^{2}-RQl^{4}A^{2}+ \left(pR_{y}+R_{z} -2\,Rp_{y} \right) l_{y}A^{2}l^{3}+\\+5 A^{2}l^{2}l_{y}R^{2} -Al^{3}l_{y}A_{z}R+R^{3} \left( 2 pl_{y}-R \right)  \Big)  \Big (Ql_{y}A^{4}l^{6}-2\,A^{2}l^{2}R \left( A^{2}l^{2}+5\,Rp \right)l_{y}^{2}-
\\-A^{2}l^{4}QR^{2}  +R  l_{y}A^{2}l^{3}( pR_{y}+R_{z}-2Rp_{y} )+6A^{2}l^{2}l_{y}R^{3}-Al^{3}l_{y}A_{z}R^{2}+R^{4} ( 2pl_{y}-R) \Big)^{-1}.
\label{eq:cc_case2_L}
\end{gathered}
\end{equation}

The next case corresponds to the vanishing of the denominator of \eqref{eq:cc_case2_L}. Consequently, we get that
 \begin{equation}
\begin{gathered}
l^{2} \left( Pl-4\,{l_{y}}^{2} \right) A^{2}+l_{y}R^{2}=0, \quad l^{2} \left( Ql^{2}-4\,Rl_{y} \right) A^{2}+R^{3}=0,\vspace{0.1cm} \\
l^{2} \left( Plp+Ql^{2}-4\,p{l_{y}}^{2}+Rl_{y}-lR_{y}\right) A^{2}+l_{y}pR^{2}=0,\vspace{0.1cm}\\
l^{2} \left( Plp^{2}+Ql^{2}p-4\,p^{2}l_{y}^{2}+2\,Rlp_{y}+6\,Rpl_{y}-lpR_{y}-R^{2}-lR_{z}\right) A^{2}-\\-2\,l_{y}A^{4}l^{4}+A_{z}Al^{3}R+pR^{2} \left( pl_{y}-R \right)=0,
\label{eq:cc_case3}
\end{gathered}
\end{equation}
while $L$ satisfies the equation
 \begin{equation}
\begin{gathered}
L^{2}-\frac{R}{A l_{y}} +\frac {l^{2}}{l_{y}}=0.
\label{eq:cc_case3_L}
\end{gathered}
\end{equation}
Finally, in the case of $l_{y}=0$ we obtain
 \begin{equation}
\begin{gathered}
l_{y}=0, \quad l_{z}^{3}-3\,p_{y}l_{z}^{2}l+3p_{y}^{2}l_{z}l^{2}-l^{3} \left( A^{2}Ql+p_{y}^{3} \right)=0,\\
pl_{z}^{3}- \left(A^{2}l^{2}+3 lpp_{y} \right) l_{z}^{2}+ \left(3l^{2}pp_{y}^{2}-A^{2}l^{3}p_{y}-Al^{3}A_{z} \right) l_{z}+\\+l^{3} \left( lp_{y}^{2}-lp_{zy}+l_{zz} \right)A^{2}+A_{z}p_{y}Al^{4}-p_{y}^{3}pl^{3}=0,
\label{eq:cc_case4}
\end{gathered}
\end{equation}
and
 \begin{equation}
\begin{gathered}
L=\frac{Al^{2}}{lp_{y}-l_{z}}.
\label{eq:cc_case4_L}
\end{gathered}
\end{equation}
We do not need to consider the case of $lp_{y}-l_{z}=0$ separately, since it results in either degeneration of transformations \eqref{eq:eq2} or reduces to subcases of \eqref{eq:cc_case2} or \eqref{eq:cc_case3}.

The above results can be summarized as follows:

\begin{theorem}
  Equation \eqref{eq:eq1a} can be transformed into \eqref{eq:eq5} if and only if one of the sets of correlations \eqref{eq:cc_case1}, \eqref{eq:cc_case2}, \eqref{eq:cc_case3} or \eqref{eq:cc_case4} holds.
  \label{th:th1}
\end{theorem}

\begin{remark}
In order to check compatibility conditions for a particular member of \eqref{eq:eq1a} one needs to calculate the values of the functions $p$ and $l$ via $g$ and $h$ with the help of the relations \eqref{eq:eq10c} taking into account that $m_{y}=g$. Then, one needs to substitute the corresponding values of the functions $p$ and $l$ into one of the sets of the compatibility conditions \eqref{eq:cc_case1}, \eqref{eq:cc_case2}, \eqref{eq:cc_case3} or \eqref{eq:cc_case4} and check whether they hold at some values of $A\neq0$ and $C$. We present a detailed algorithm for verifying compatibility conditions at the beginning of the next section.
\end{remark}

As an immediate consequence of Theorem \ref{th:th1} we get
\begin{corollary}
If one of the sets of correlations \eqref{eq:cc_case1}, \eqref{eq:cc_case2}, \eqref{eq:cc_case3} or \eqref{eq:cc_case4} holds then equation \eqref{eq:eq1a} admits the following first integral
 \begin{equation}
 I=A^{-2\rho}\left(2y_{z}+2p-\frac{\beta-\rho}{\beta}AL\right)^{\rho+\beta}\left(2y_{z}+2p-\frac{\beta+\rho}{\beta}AL\right)^{\rho-\beta},
 \label{eq:fi_general}
 \end{equation}
where $p$ is given in \eqref{eq:eq10c}, $\rho=\sqrt{\beta^{2}-4\alpha}\neq0$ and $L=\beta F$. If $\rho=0$ then the first integral is
 \begin{equation}
 I=\frac{2y_{z}+2p-AL}{A}\exp\left\{\frac{AL}{2y_{z}+2p-AL}\right\}.
 \end{equation}
\label{cr:cr1}
\end{corollary}

It is interesting to understand when transformations \eqref{eq:eq2} keep first integral \eqref{eq:eq6a} autonomous. One can show that this is true if and only if $G_{z}=F_{z}=0$. As a consequence, we have that the following statement holds
\begin{corollary}
Equation of the form
\begin{equation*}
  y_{zz}+g(y)y_{z}+h(y)=0,
\end{equation*}
is integrable with the first integral
\begin{equation*}
 I=\left(2\beta y_{z}+(\rho+\beta)(m+\mu)\right)^{\rho+\beta}\left(2\beta y_{z}+(\rho-\beta)(m+\mu)\right)^{\rho-\beta},
\end{equation*}
if
\begin{equation*}
  \beta^{2}(hg_{y}-gh_{y})+\alpha g^{3}=0,
\end{equation*}
where $m_{y}=g$.

\end{corollary}

Let us also consider the case when transcendental first integral \eqref{eq:fi_general} becomes a rational one. One can show that this is true if the following relation holds $4\alpha=(1-r^{2})\beta$, where $r\neq0$ is a rational number. As a consequence, we have that
\begin{corollary}
If one of the sets of correlations \eqref{eq:cc_case1}, \eqref{eq:cc_case2}, \eqref{eq:cc_case3} or \eqref{eq:cc_case4} holds and
 \begin{equation}
 4\alpha=(1-r^{2})\beta, \quad r=\frac{n}{k}, \quad k,n \in \mathbb{Z}\setminus\{0\},
  \end{equation}
 then equation \eqref{eq:eq1a} admits the following rational first integral
 \begin{equation}
 I=A^{-2n}\left(2y_{z}+2p-(1-r)AL\right)^{n+k}\left(2y_{z}+2p-(1+r)AL\right)^{n-k},
 \label{eq:fi_rational}
 \end{equation}
where $p$ is given in \eqref{eq:eq10c} and $L=\beta F$.
\end{corollary}

Thus, in this section we have explicitly find correlations on functions $g$ and $h$ that give us the linearization criterion for \eqref{eq:eq1a} via generalized Sundman transformations. We have also showed that once an equation from \eqref{eq:eq1} is linearizable it possesses a certain first integral. Moreover, we have isolated linearizable families of equations that admits an autonomous first integral or a rational one.

\section{Examples}

In this section we provide several new examples of linearizable equations of form \eqref{eq:eq1}. First, we demonstrate that there are indeed equations from family \eqref{eq:eq1} with coefficient satisfying conditions from Theorem \ref{th:th1}, but which cannot be linearized via \eqref{eq:eq2} with $F_{z}=0$. Then, we provide several example of both non-autonomous and autonomous nonlinear oscillators including generalizations of the Duffing and Van der Pol oscillators, that can be linearized via \eqref{eq:eq2} with $F_{z}\neq0$.

Let us present an algorithm for verifying that a particular member of \eqref{eq:eq1a} can be linearized with the help of \eqref{eq:eq2}. It consists of the following three steps.
First, using \eqref{eq:eq10c} and taking into account that $m_{y}=g$ we calculate the values of the functions $p$ and $l$ via $g$ and $h$. Second, we substitute the corresponding values of the functions $p$ and $l$ into one of compatibility conditions \eqref{eq:cc_case1}, \eqref{eq:cc_case2}, \eqref{eq:cc_case3} or \eqref{eq:cc_case4}. As a result of this substitution, we obtain polynomials in $y$, whose coefficients are functions of $z$. Equating coefficients of these polynomials to zero, we get a system of equations for the functions $A$  and $C$. If this system is satisfied for any values of $A\neq0$ and $C$, then the corresponding equation from \eqref{eq:eq1a} is linearizable. Third, if one of the sets of the compatibility conditions is satisfied, we calculate the value of $L$ via one of the relations \eqref{eq:cc_case1_L}, \eqref{eq:cc_case2_L}, \eqref{eq:cc_case3_L}, \eqref{eq:cc_case4_L} and then it is easy to find the explicit form of the linearizing transformations with the help of \eqref{eq:eq9} and \eqref{eq:eq10c}.

\textbf{Example 1}. Let us consider the following equation from family \eqref{eq:eq1a}
\begin{equation}\label{eq:ex3_1}
y_{zz}+(\beta e^{-\delta z}y-2\delta)y_{z}+\frac{e^{-\delta z}}{2}y^{2}(\alpha e^{-\delta z}y-2\beta\delta)=0.
\end{equation}
In order to check that this equation can be linearized via \eqref{eq:eq2} we use the algorithm presented above. With the help of \eqref{eq:eq10c} we find that $p=\beta e^{-\delta z}/2 y^{2}-\delta y-\beta\delta^{2}e^{\delta z}/\alpha$ and $l=2\beta^{2}e^{-2\delta z}y(\alpha e^{-2\delta z}y^{2}-2\delta^{2})/\alpha$. Substituting these values of $p$ and $l$ into \eqref{eq:cc_case1} and equating coefficients at the same powers of $y$ we find that $A=e^{\delta z}/2$ and $C=-\beta\delta^{2}e^{\delta z}/\alpha$ and from \eqref{eq:cc_case1_L}, \eqref{eq:eq9} and \eqref{eq:eq10c} we get that $L=\beta (e^{-2\delta z}y^{2}-2\delta^{2}/\alpha)$, $F=e^{-2\delta z}y^{2}-2\delta^{2}/\alpha$ and $G={\rm e}^{-\delta z}y$.  As a result, we have that \eqref{eq:ex3_1} can be linearized via \eqref{eq:eq2} and its general solution can be presented in the following parametric form
\begin{equation}\label{eq:ex3_2}
y=\pm e^{\delta z}\left(w+\frac{2\delta^{2}}{\alpha}\right)^{1/2}, \quad
z=\pm \bigintssss \frac{d\zeta}{\left(w+\frac{2\delta^{2}}{\alpha}\right)^{1/2}},
\end{equation}
where $w$ is the general solution of \eqref{eq:eq5}.

From Corollary \ref{cr:cr1} it follows that \eqref{eq:ex3_1} possesses the first integral
\begin{equation}\label{eq:ex3_3}
\begin{gathered}
I={\rm e}^{-2\rho\delta z}\left(2y_{z}-2\delta y+\beta e^{-\delta z}y^{2}-\frac{2\beta\delta^{2}}{\alpha}e^{\delta z}-\frac{e^{\delta z}}{2\alpha}(\beta-\rho)(\alpha e^{-2\delta z }y^{2}-2\delta^{2})\right)^{\rho+\beta}\\
\left(2y_{z}-2\delta y+\beta e^{-\delta z}y^{2}-\frac{2\beta\delta^{2}}{\alpha}e^{\delta z}-\frac{e^{\delta z}}{2\alpha}(\beta+\rho)(\alpha e^{-2\delta z }y^{2}-2\delta^{2})\right)^{\rho-\beta},
\end{gathered}
\end{equation}
if $\rho\neq0$ and the first integral
\begin{equation}\label{eq:ex3_3a}
\begin{gathered}
I=e^{\delta z}\left(4y_{z}-4\delta y+\beta \frac{y^{2}}{e^{\delta z}}-\frac{8\delta^{2}}{\beta}e^{\delta z}\right)
\exp\left\{\frac{8\delta^{2}e^{2\delta z}-\beta^{2}y^{2}}{8\delta^{2}e^{2\delta z}-\beta^{2}y^{2}+4\beta \delta  e^{\delta z}y-4 \beta e^{\delta z} y_{z}}\right\},
\end{gathered}
\end{equation}
if $\rho=0$.

Equation \eqref{eq:ex3_1} can be considered as a non-autonomous generalization of the damped Duffing oscillator. Notice that one can show that equation \eqref{eq:ex3_1} cannot be linearized via \eqref{eq:eq2} with $F_{z}=0$ and possesses only one Lie point symmetry. Therefore, \eqref{eq:ex3_1} provides an example of an equation that cannot be both integrated with the help of the classical Lie approach and linearized with the help of the restricted case of transformations \eqref{eq:eq2}.

\begin{figure}[!t]
\center
\includegraphics[width=0.9\textwidth]{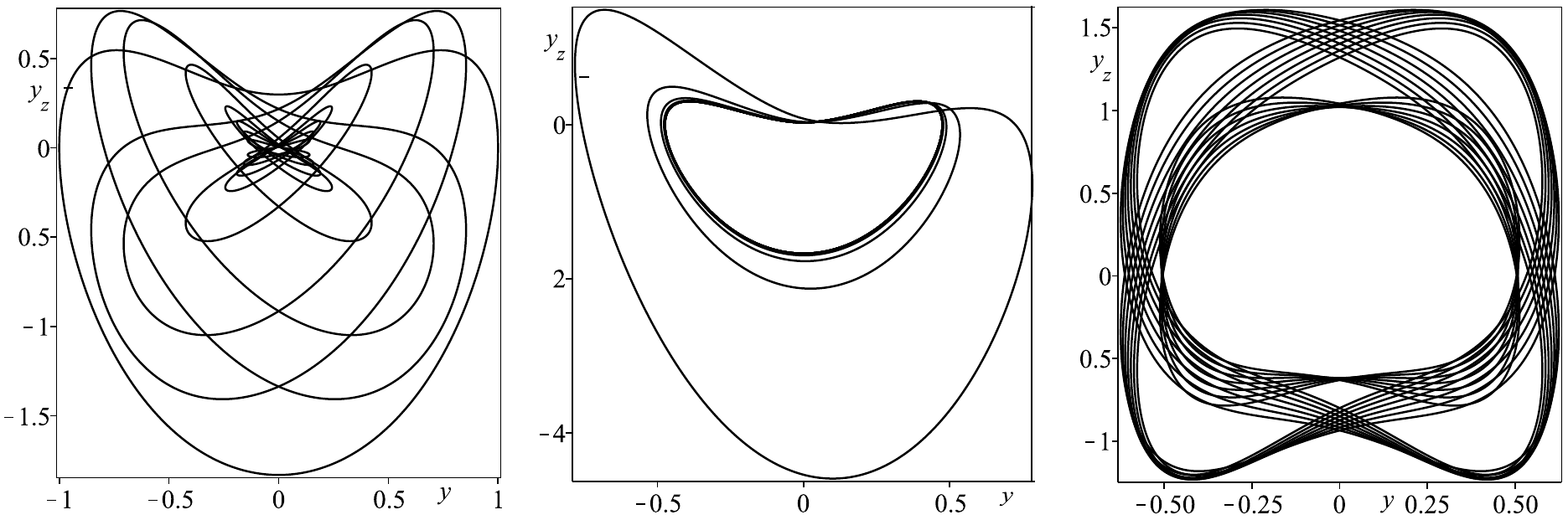} 
\caption{Projections of \eqref{eq:ex2_2} on the plane $z=c$ for different values of $c$: $s(z)=\sin z$, $\alpha=-\beta=1$ (left figure); $s(z)={\rm e}^{-2z}$, $\alpha=10$, $\beta=-5$ (middle figure); $s(z)=\sin z \cos (\pi z)$, $\alpha=10$, $\beta=1$ (right figure).   }
\label{fig1}
\end{figure}

\textbf{Example 2.}
Consider a family of parametrically forced Duffing oscillators with linear damping
 \begin{equation}
 y_{zz}+(b_{1} y+b_{0}(z))y_{z}+a_{3}y^{3}+a_{2}(z) y^{2}+a_{1}(z) y=0,
 \label{eq:ex2_1a}
\end{equation}
where $b_{1}\neq0$ and $a_{3}\neq0$ are certain parameters and $b_{0}$, $a_{2}$ and $a_{1}$ are certain functions of $z$. Now we need to check whether coefficients of \eqref{eq:ex2_1a} satisfy one of the sets of the compatibility conditions. For the sake of simplicity, we assume that $C=0$. The case of $C\neq0$ can be treated in the same way.

According to the algorithm presented above, at the first step we find that
\begin{equation}
\begin{gathered}
\label{eq:ex2_1b}
  p=\frac{A_{z}}{A}y+\frac{b_{2}y^{2}}{2}+b_{0}y, \\
  l=\frac{\beta^{2}}{2\alpha A^{4}}\left[2(a_{3}y^{3}+a_{2}y^{2}-b_{0,z}y+a_{1}y)A^{2}+(b_{2}y^{2}+2b_{0}y)A_{z}-2AA_{zz}y+4A_{z}^{2}y\right].
  \end{gathered}
\end{equation}
Substituting \eqref{eq:ex2_1b} into \eqref{eq:cc_case1} and collecting coefficients at the same powers of $y$, we find that if
\begin{equation}\label{eq:ex2_1c}
  b_{2}=2\beta, \quad b_{0}=3s, \quad a_{3}=2\alpha, \quad a_{2}=2\beta s, \quad a_{1}=2s^{2}+s_{z},\quad A(z)={\rm e}^{-2 \int s(z)dz},
\end{equation}
then conditions \eqref{eq:cc_case1} are satisfied. Here $s=s(z)$ is an arbitrary function. As a consequence, with the help of \eqref{eq:cc_case1_L}, \eqref{eq:eq10c} and \eqref{eq:eq9},  we get that
$F={\rm e}^{2\int s dz}y^{2}$ and $G=2y$.

\begin{figure}[!t]
\center
\includegraphics[width=0.3\textwidth]{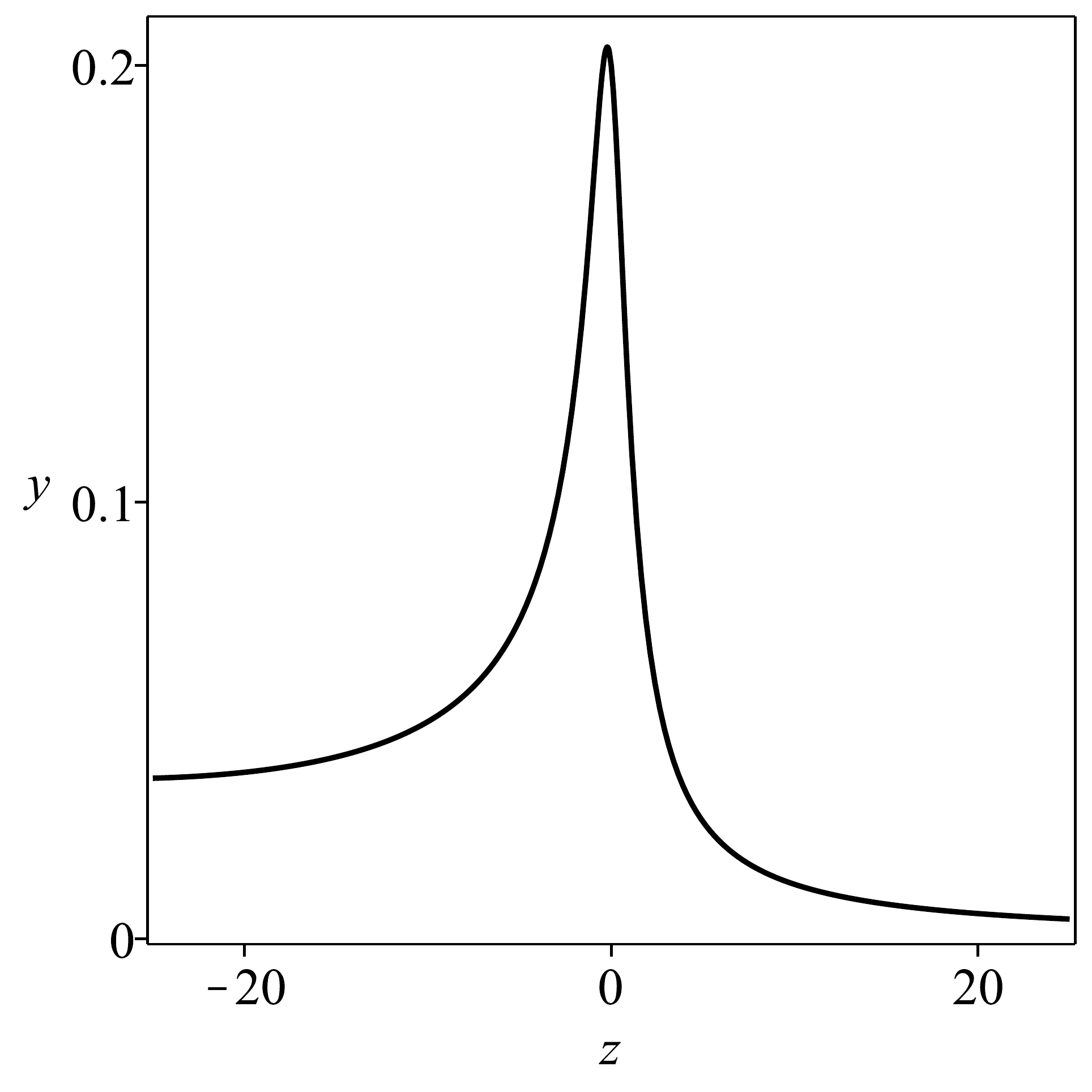} 
\includegraphics[width=0.3\textwidth]{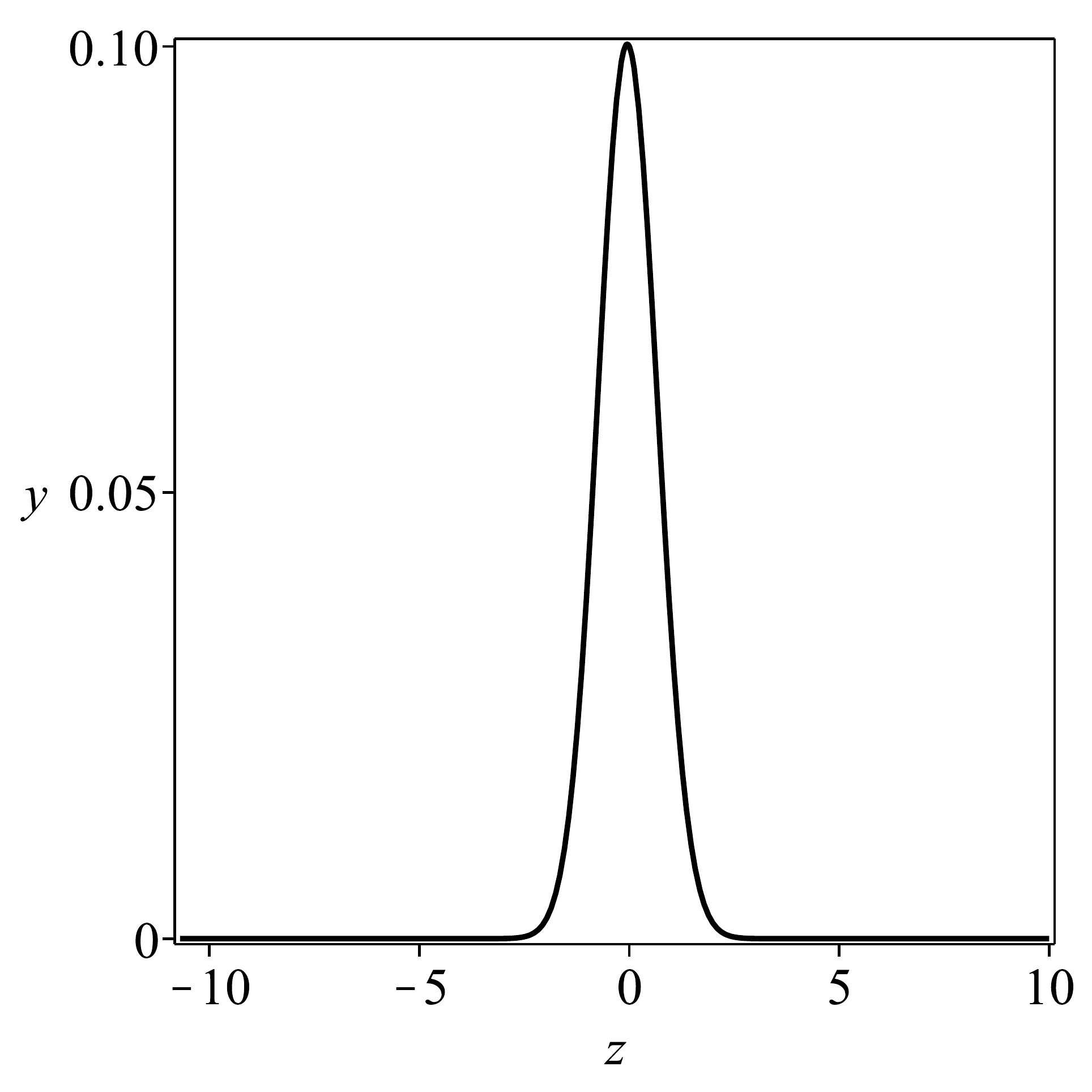}
\includegraphics[width=0.3\textwidth]{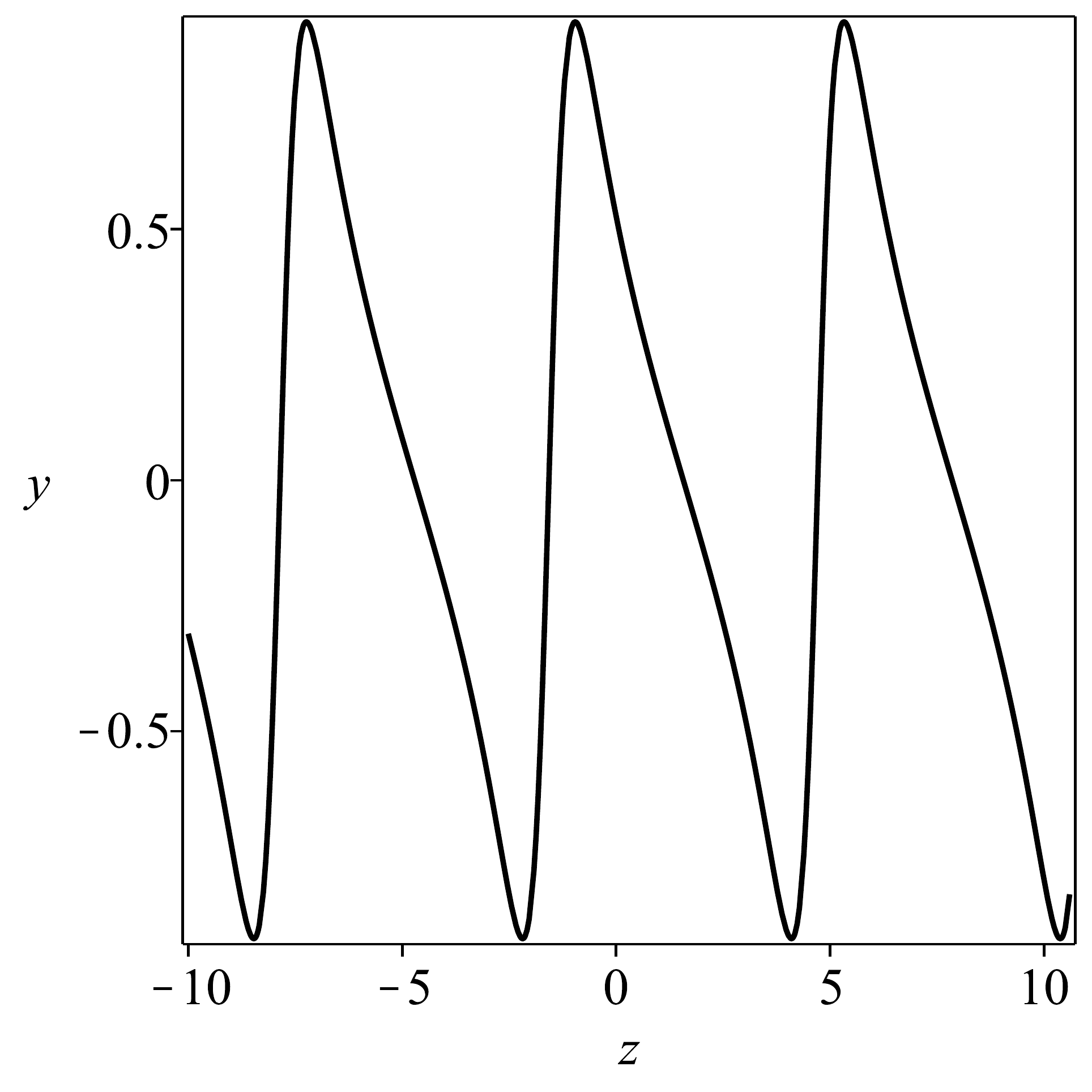}
\caption{Plots of one parametric families of solutions of \eqref{eq:ex2_1} for different forcing functions: $s(z)=z/(z^{2}+1)$, $\alpha=4$, $\beta=5$ (left figure); $s(z)=2z$, $\alpha=4$, $\beta=5$ (middle figure); $s(z)=\tan z$, $\alpha=3.8$, $\beta=4$ (right figure).}
\label{fig2}
\end{figure}

As a result, we have that the equation
 \begin{equation}
 y_{zz}+(2\beta y+3s)y_{z}+2\alpha y^{3}+2 \beta s y^{2}+(2s^{2}+s_{z}) y=0,
 \label{eq:ex2_1}
\end{equation}
can be linearized with the help of \eqref{eq:eq2}.

From Corollary \ref{cr:cr1} it follows that \eqref{eq:ex2_1} has the following first integral if $\rho\neq0$
\begin{equation}
 I={\rm e}^{4\rho\int s dz} \left(2y_{z}+2sy+(\beta+\rho)y^{2}\right)^{\rho+\beta}\left(2y_{z}+2sy+(\beta-\rho)y^{2}\right)^{\rho-\beta},
  \label{eq:ex2_2}
\end{equation}
and if $\rho=0$ this first integral is
\begin{equation}
 I={\rm e}^{2\int sdz}\left(2y_{z}+2sy+\beta y^{2}\right)\exp\left\{\frac{\beta y^{3}}{2y_{z}+2s y+\beta y^{2}}\right\}.
  \label{eq:ex2_2a}
\end{equation}

The general solution of \eqref{eq:ex2_1} can be presented as follows
\begin{equation}
y=\pm\sqrt{w}{\rm e}^{\int s dz},
 \label{eq:ex2_3}
\end{equation}
where $z$ is given by
\begin{equation}
\pm \bigintssss \frac{d\zeta}{2\sqrt{w}}=\bigintssss {\rm e}^{-\int s dz}dz,
 \label{eq:ex2_4}
\end{equation}
and $w$ is the general solution of \eqref{eq:eq5}.

Let us discuss some properties of solutions of \eqref{eq:ex2_1}. In Fig.\ref{fig1} we demonstrate projections of \eqref{eq:ex2_2} on the plane $z=\mbox{const}$ for different values the forcing function $s$ and other parameters. Notice that for the left figure the integration constant corresponds to $y(0)=1, y_{z}(0)=0$ and for the other cases the integration constant corresponds to $y(0)=1/2, y_{z}(0)=1/5$. One can see that equation \eqref{eq:ex2_1} has various types of periodic solutions even if the forcing function is not periodic. Furthermore, one can show by varying the integration constant that these periodic trajectories are not isolated in the phase space, namely, they are not limit cycles.

One-parametric families of solutions of \eqref{eq:ex2_1} can be easily obtained from \eqref{eq:ex2_2} and \eqref{eq:ex2_2a} as follows
\begin{equation}
y=\frac{2\exp\{-\int s(z)dz\}}{(\beta\pm\rho)\int \exp\{-\int s(z)dz\}dz+C_{1} },
 \label{eq:ex2_5}
\end{equation}
where $C_{1}$ is an arbitrary constant. We demonstrate plots of \eqref{eq:ex2_5} for different forcing functions and values of parameters in Fig.\ref{fig2}. One can see that depending on the forcing functions these solutions may be solitary or periodic waves.

\begin{figure}[!t]
\center
\includegraphics[width=0.4\textwidth,height=0.3\textheight]{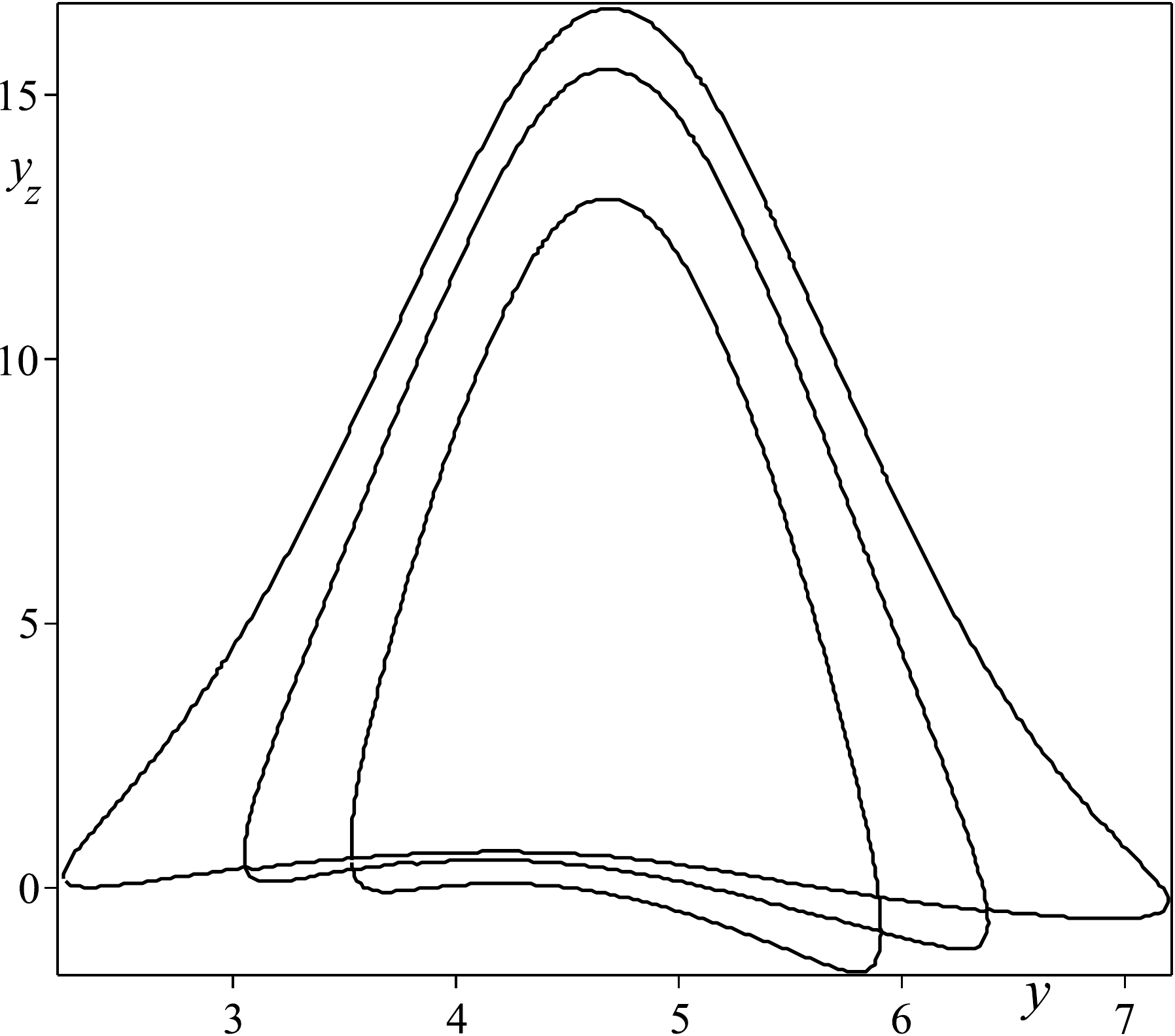} 
\includegraphics[width=0.4\textwidth,height=0.3\textheight]{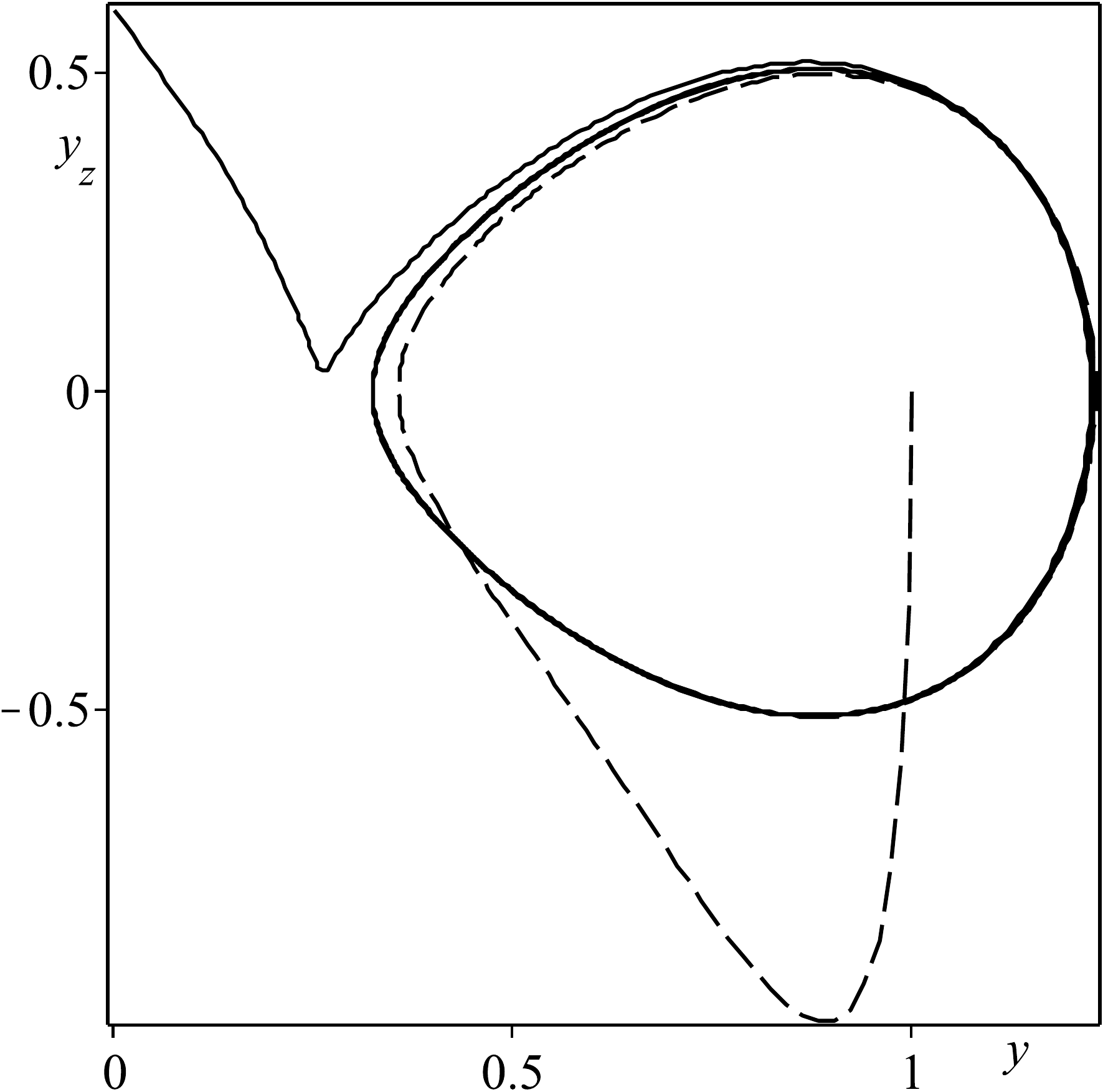}
\caption{Projections of \eqref{eq:ex4_2} on the plane $y=c$ for different values of $c$ (left figure)  and two numerical solutions of the Cauchy problem for \eqref{eq:ex4_1} (right figure) at $\alpha=4$, $\beta=5$, $\mu=-1$ and $s(z)={\rm e}^{\sin z}$ (right figure).}
\label{fig3}
\end{figure}

\textbf{Example 3.} Let us consider the following family of non-autonomous nonlinear oscillators
\begin{equation}
y_{zz}+\left(b_{2} y^{2}+b_{0}\right)y_{z}+a_{5}y^{5}+a_{3}y^{3}+a_{2} y^{2}+a_{1}y=0,
 \label{eq:ex4_1a}
\end{equation}
where $b_{i}=b_{i}(z),\,i=1,2$ and $a_{j}=a_{j}(z),j=1,2,3,5$ are some functions and $b_{2},\,a_{5}\not\equiv0$. Equation \eqref{eq:ex4_1a} can be considered as a parametrically forced $\phi^{6}$--Van der Pol oscillator or as a parametrically forced extended Duffing--Van der Pol system (see, e.g. \cite{Siewe2004,Yu2009}).

Now we find a case of \eqref{eq:ex4_1a}, whose coefficients satisfy \eqref{eq:cc_case1}. First, we compute the values of $l$ and $p$
\begin{equation}
\begin{gathered}
p=\frac{A_{z}y}{A}+\frac{b_{2}y^{3}}{3}+b_{0}y+C, \quad l=\frac{\beta^{2}}{3\alpha A^{4}} \Big[(b_{2}y^{3}+3b_{0}y+3C)AA_{z}+\\
+ 3(2A_{z}^{2}-AA_{zz})y+(3a_{5}y^{5}+(3a_{3}-b_{2,z})y^{3}+3a_{2}y^{2}+3(a_{1}-b_{0,z})y-3C_{z})A^{2}\Big].
 \label{eq:ex4_1b}
 \end{gathered}
\end{equation}
Second, we substitute these values into \eqref{eq:cc_case1} and equate coefficients at the same powers of $y$. As a consequence, we obtain that
\begin{equation}
\begin{gathered}
b_{5}=3\beta s, \quad b_{0}=\frac{5s_{z}}{3s}, \quad a_{5}=3\alpha s^{2}, \quad a_{3}=2\beta s_{z}, \quad a_{2}=3\mu,\\
a_{1}=\frac{2s_{zz}}{3s}, \quad A=1/s, \quad C=\beta \mu /(\alpha s).
 \label{eq:ex4_1c}
 \end{gathered}
\end{equation}
Here $s(z)\not\equiv 0$ is an arbitrary sufficiently smooth function and $\mu$ is an arbitrary parameter.

Third, using \eqref{eq:ex4_1c}, \eqref{eq:cc_case1_L}, \eqref{eq:eq9} and \eqref{eq:eq10c} we get that the equation
\begin{equation}
y_{zz}+\left(3\beta s y^{2}+\frac{5s_{z}}{3s}\right)y_{z}+3\alpha s^{2}y^{5}+2\beta s_{z}y^{3}+3\mu y^{2}+\frac{2s_{zz}}{3s}y=0,
 \label{eq:ex4_1}
\end{equation}
can be linearized via transformations \eqref{eq:eq2} with
\begin{equation}
F=s^{2}y^{3}+\frac{\mu}{\alpha}, \quad G=3 s y^{2}.
 \label{eq:ex4_1e}
\end{equation}
Notice that the general solution of \eqref{eq:ex4_1} in a nonlocal form can be obtained by inverting \eqref{eq:eq2} with \eqref{eq:ex4_1e} as follows
\begin{equation}\label{eq:ex4_1d}
  y^{3}=\frac{1}{s^{2}}\left(w-\frac{\mu}{\alpha}\right), \quad \int s dz=\int \frac{d\zeta}{3y^{2}},
\end{equation}
where $w$ is the general solution of \eqref{eq:eq5}.

With the help of \ref{cr:cr1} we find that \eqref{eq:ex4_1} possesses a first integral
\begin{equation}
I=\left(6\alpha s y_{z}+4\alpha y s_{z}+3(\beta+\rho)(\alpha s^{2}y^{3}+\mu)\right)^{\rho+\beta}\left(6\alpha s y_{z}+4\alpha y s_{z}+3(\beta-\rho)(\alpha s^{2}y^{3}+\mu)\right)^{\rho-\beta},
 \label{eq:ex4_2}
\end{equation}
if $\rho\neq0$, and a first integral
\begin{equation}
I=(6\beta s y_{z}+4\beta s_{z}y+3\beta^{2} s^{2}y^{3}+12\mu)\exp\left\{\frac{3\beta^{2}s^{2}y^{3}+12\mu}{6\beta s y_{z}+4 \beta s_{z}y+3\beta^{2}s^{2}y^{3}+12\mu}\right\},
 \label{eq:ex4_3}
\end{equation}
if $\rho=0$.

Now we discuss some properties of integrals \eqref{eq:ex4_2} and \eqref{eq:ex4_3}. In the left part of Fig. \ref{fig3} we demonstrate projections of \eqref{eq:ex4_2} at certain values of the parameters on the $y=\rm{const}$ plane for different values of this constant. One can see that there are periodic trajectories admitted by \eqref{eq:ex4_1}. We argue that these periodic trajectories are limit cycles. To support this claim in the right part of Fig. \ref{fig3} we demonstrate the results of numerical solution of the Cauchy problem for \eqref{eq:ex4_1}. We see that nearby trajectories in the phase space converge to a certain closed trajectory, and, thus, there is indeed a limit cycle in \eqref{eq:ex4_1}. One can also observe a similar situation for first integral \eqref{eq:ex4_3}. Finally, if we assume that the forcing function $s(z)$ is periodic and has no zeros on the real line, one can again find a limit cycle in \eqref{eq:ex4_1}.

\textbf{Example 4.} Now we consider the following equation from family \eqref{eq:eq1}
\begin{equation}
y_{zz}+\frac{y_{z}^{2}}{y}+\left(\mu+\frac{\beta}{y^{3}}\right)y_{z}-\mu^{2}y+\frac{\beta\mu}{y^{2}}-\frac{\alpha}{y^{5}}=0,
 \label{eq:ex5_1}
\end{equation}
where $\mu\neq0$ is an arbitrary parameter. If we transform \eqref{eq:ex5_1} into its canonical form via $y\rightarrow \sqrt{2y}$, one can verify with the help of the above proposed algorithm, that the coefficients of the corresponding equation of type \eqref{eq:eq1a} satisfy conditions \eqref{eq:cc_case1}. Indeed, in this case we have that $p=2\mu y-\beta/\sqrt{2y}$, $l=-\beta^{2}{\rm e}^{-2\mu z}/(4y^{2})$ and $A=-{\rm e}^{-\mu z}$ satisfy \eqref{eq:cc_case1} and $L=\beta {\rm e}^{-\mu z}/\sqrt{2y}$. Consequently, equation \eqref{eq:ex5_1} can be linearized via \eqref{eq:eq2} with
\begin{equation}
F=\frac{{\rm e}^{-\mu z}}{y}, \quad G=\frac{1}{y^{3}}.
\end{equation}
The general solution of \eqref{eq:ex5_1} can be expressed in the parametric form as follows
\begin{equation}
y=\frac{{\rm e}^{-\mu z}}{w}, \quad z=\frac{1}{3\mu}\ln \left\{3\mu \int\frac{d\zeta}{w^{3}} \right\}.
\end{equation}
With the help of Corollary \ref{cr:cr1} we find that \eqref{eq:ex5_1} has the first integral
\begin{equation}
I=\left(2e^{-\mu z}y(y_{z}+\mu y)-(\beta+\rho)\frac{e^{-\mu z}}{y}\right)^{\rho+\beta}\left(2e^{-\mu z}y(y_{z}+\mu y)-(\beta-\rho)\frac{e^{-\mu z}}{y}\right)^{\rho-\beta},
\end{equation}
if $\rho\neq0$ and the first integral
\begin{equation}
I=e^{-\mu z}\left(2y(y_{z}+\mu y)-\frac{\beta }{y}\right)\exp\left\{\frac{\beta}{\beta-2\mu y^{3}-2 y^{2}y_{z}}\right\},
\end{equation}
if $\rho=0$. Notice that equation \eqref{eq:ex5_1} provides an example of an autonomous nonlinear oscillator from family \eqref{eq:eq1}, i.e. an equation with quadratic nonlinearity with respect to the first derivative, that can be linearized via \eqref{eq:eq2} with $F_{z}\neq0$.



In this section we have provided several examples of linearizable equations from family \eqref{eq:eq1} that can be transformed into \eqref{eq:eq3} via \eqref{eq:eq2} only if $F_{z}=0$. We have also showed that the corresponding first integrals allows us to find periodic trajectories. including limit cycles, admitted by the considered nonlinear oscillators.

\section{Conclusion}
In this work we have considered family \eqref{eq:eq1} of nonlinear second order ordinary differential equations. We have studied the complete linearization problem for this family of equations via the generalized Sundman transformations and obtained linearizability conditions in the explicit form. We have also shown that each linearizable equation from \eqref{eq:eq1} admits a certain transcendental first integral. As a consequence, we classify all equations of form \eqref{eq:eq1} that possess this transcendental first integral. We have also separated families of equations with autonomous and rational first integral. We have provided several nontrivial examples of applications of the linearizing transformations including generalizations of the Duffing and Van der Pol oscillators. In particular, we have demonstrated that our approach can be used for finding centers and limit cycles admitted by equations from the considered family.

\section{Acknowledgments}
This research was supported by Russian Science Foundation grant No. 19-71-10003. Numerical calculations in Section 3 were supported by  Russian Science Foundation grant  No. 19-71-10048.

\end{document}